\def\figuresize{3.4in}
\def\figuresizeThird{2.33in}
\def\complexNumbers{\mathbb{C}}
\def\realNumbers{\mathbb{R}}
\def\integers{\mathbb{Z}}

\def\expectationOperator[#1][#2]{{\mathbb{E}_{#2}}\left[#1\right]}
\def\probability[#1]{\textrm{Pr}\left({#1}\right)}
\def\complexGaussian[#1][#2]{\mathcal{CN}({#1,#2})}
\def\gaussianDist[#1][#2]{\mathcal{N}({#1,#2})}
\def\uniformDist[#1][#2]{\mathcal{U}({#1,#2})}
\def\indicatorFunction[#1]{\mathbb{I}\left[{#1}\right]}
\def\numberOfEdgeDevices{K}
\def\centroid[#1][#2]{{\textrm{\textbf{c}}}_{#1}^{(#2)}}
\def\centroidNoise[#1][#2]{{\textrm{\textbf{n}}}_{#1}^{(#2)}}
\def\varianceCentroid{\sigma^2_{\textrm{c}}}
\def\minimumCardinality{S_{\textrm{min}}}
\def\dataset[#1]{\mathcal{D}_{#1}}
\def\partition[#1]{\mathcal{S}_{#1}}
\def\partitionIterations[#1][#2]{\mathcal{S}_{#1}^{(#2)}}
\def\meanPartition[#1]{\boldsymbol{\mu}_{#1}}
\def\completeData{\mathcal{D}}
\def\dataSample[#1]{{\textrm{\textbf{d}}}_{#1}}
\def\learningRate{\mu}
\def\localDataSetGivenCentroid[#1][#2]{\mathcal{D}_{#1}^{(#2)}}
\def\localUpdateVector[#1][#2]{\Delta{\textrm{\textbf{c}}}_{#1}^{(#2)}}
\def\lengthOfDataSample{L}
\def\indexCommunicationRound{n}
\def\indexED{k}
\def\indexCentroid{c}
\def\indexCentroidAnother{c'}
\def\numberOfCentroids{C}
\def\lossFunction[#1]{f(#1)}
\def\indexGradient{q}
\def\communicationRounds{N}

\def\noiseVariance{\sigma_{\rm n}^2}

\def\spectralEfficiency{r_{\text{bits}}}
\def\compressionRatio{r_{\text{compression}}}
\def\numberOfBits{N_{\text{bits}}}

\def\base{\beta}
\def\indexDigit{d}

\def\numberOfDigits{D}
\def\normalizationDigit{\xi}
\def\valueMaximum{v_{\rm max}}

\def\representationInBase[#1][#2]{\left(#2\right)}
\def\digitAveraged[#1][#2]{\sigma_{{#1},{#2}}^{(\indexCommunicationRound)}}
\def\digitAveragedEst[#1][#2]{\hat{\sigma}_{{#1},{#2}}^{(\indexCommunicationRound)}}
\def\digitGeneral[#1]{s_{#1}}
\def\digit[#1][#2]{\eta_{{#1},{#2}}^{(\indexCommunicationRound)}}
\def\digitStandart[#1]{b_{#1}}

\def\symbolWithoutZero[#1]{b_{#1}}
\def\resourceSet[#1]{\mathbb{T}_{#1}}

\def\symbolSet[#1]{\mathbb{S}_#1}
\def\symbol[#1]{s_{#1}}
\def\indexSymbol{j}
\def\decoder{f_{\text{dec},\base}}
\def\encoder{f_{\text{enc},\base}}

\def\aRealValueClamped{v'}

\def\numberOFEDsForOptionGeneralDetector[#1]{\hat{K}_{\indexGradient,\indexDigit,#1}}
\def\meanGradientEleEstimate[#1][#2]{{\hat{v}}^{(#1)}_{#2}}
\def\meanGradientEle[#1][#2]{{v}^{(#1)}_{#2}}

\def\receivedSymbolAtSubcarrier[#1]{y_{#1}^{(\indexCommunicationRound)}}
\def\transmittedSymbolAtSubcarrier[#1]{x_{#1}^{(\indexCommunicationRound)}}
\def\indexSubcarrier{l}
\def\channelAtSubcarrier[#1]{h_{#1}^{(\indexCommunicationRound)}}
\def\noiseAtSubcarrier[#1]{w_{#1}^{(\indexCommunicationRound)}}
\def\randomSymbolAtSubcarrier[#1]{r_{#1}^{(\indexCommunicationRound)}}
\def\voteInTime[#1][#2]{m_{#1,#2}}
\def\voteInFrequency[#1][#2]{l_{#1,#2}}
\def\symbolEnergy{E_{\rm s}}
\def\numberOFEDsForOptionGeneral[#1]{K_{\indexGradient,\indexDigit,#1}}
\def\SNR{{\tt{SNR}}}

\def\metric[#1][#2]{m_{#1}^{(#2)}}
\def\metricVector[#1]{\textbf{m}^{(#1)}}
\def\metricFactor{\alpha}
\def\indicesNullSet{\mathbb{E}}

\documentclass[conference]{IEEEtran}

\usepackage{multicol}
\usepackage{lipsum}
\usepackage{mathtools}
\usepackage{amsthm}
\usepackage{acronym}
\usepackage{stfloats}
\usepackage{amsfonts}
\usepackage{acronym}
\usepackage[noadjust]{cite}
\usepackage{multirow}
\usepackage{bm}
\usepackage{amsmath,amssymb}
\usepackage{graphicx}
\usepackage{epstopdf}
\usepackage[table,xcdraw]{xcolor}
\usepackage[caption=false,font=footnotesize]{subfig}
\usepackage[geometry]{ifsym}
\usepackage{array}
\usepackage[utf8]{inputenc}
\usepackage[T1]{fontenc}
\usepackage{pifont}
\usepackage{tabularray}
\usepackage{lipsum}
\usepackage{algpseudocode}
\usepackage[ruled,vlined]{algorithm2e}
\def\BibTeX{{\rm B\kern-.05em{\sc i\kern-.025em b}\kern-.08em
		T\kern-.1667em\lower.7ex\hbox{E}\kern-.125emX}}

\let\norm\undefined 
\DeclarePairedDelimiter\norm{\lVert}{\rVert}
\newcommand\mydots{\hbox to 1em{.\hss.\hss.}}

\DeclarePairedDelimiter\floor{\lfloor}{\rfloor}

\newcommand{\vectorization}[1]{\mathrm{vec}(#1)}

\makeatletter
\newif\ifAC@uppercase@first%
\def\Aclp#1{\AC@uppercase@firsttrue\aclp{#1}\AC@uppercase@firstfalse}%
\def\AC@aclp#1{%
	\ifcsname fn@#1@PL\endcsname%
	\ifAC@uppercase@first%
	\expandafter\expandafter\expandafter\MakeUppercase\csname fn@#1@PL\endcsname%
	\else%
	\csname fn@#1@PL\endcsname%
	\fi%
	\else%
	\AC@acl{#1}s%
	\fi%
}%
\def\Acp#1{\AC@uppercase@firsttrue\acp{#1}\AC@uppercase@firstfalse}%
\def\AC@acp#1{%
	\ifcsname fn@#1@PL\endcsname%
	\ifAC@uppercase@first%
	\expandafter\expandafter\expandafter\MakeUppercase\csname fn@#1@PL\endcsname%
	\else%
	\csname fn@#1@PL\endcsname%
	\fi%
	\else%
	\AC@ac{#1}s%
	\fi%
}%
\def\Acfp#1{\AC@uppercase@firsttrue\acfp{#1}\AC@uppercase@firstfalse}%
\def\AC@acfp#1{%
	\ifcsname fn@#1@PL\endcsname%
	\ifAC@uppercase@first%
	\expandafter\expandafter\expandafter\MakeUppercase\csname fn@#1@PL\endcsname%
	\else%
	\csname fn@#1@PL\endcsname%
	\fi%
	\else%
	\AC@acf{#1}s%
	\fi%
}%
\def\Acsp#1{\AC@uppercase@firsttrue\acsp{#1}\AC@uppercase@firstfalse}%
\def\AC@acsp#1{%
	\ifcsname fn@#1@PL\endcsname%
	\ifAC@uppercase@first%
	\expandafter\expandafter\expandafter\MakeUppercase\csname fn@#1@PL\endcsname%
	\else%
	\csname fn@#1@PL\endcsname%
	\fi%
	\else%
	\AC@acs{#1}s%
	\fi%
}%
\edef\AC@uppercase@write{\string\ifAC@uppercase@first\string\expandafter\string\MakeUppercase\string\fi\space}%
\def\AC@acrodef#1[#2]#3{%
	\@bsphack%
	\protected@write\@auxout{}{%
		\string\newacro{#1}[#2]{\AC@uppercase@write #3}%
	}\@esphack%
}%
\def\Acl#1{\AC@uppercase@firsttrue\acl{#1}\AC@uppercase@firstfalse}
\def\Acf#1{\AC@uppercase@firsttrue\acf{#1}\AC@uppercase@firstfalse}
\def\Ac#1{\AC@uppercase@firsttrue\ac{#1}\AC@uppercase@firstfalse}
\def\Acs#1{\AC@uppercase@firsttrue\acs{#1}\AC@uppercase@firstfalse}

\acrodef{WSN}{wireless sensor network}
\acrodef{USRP}{universal software radio peripheral}
\acrodef{SN}{sensor node}
\acrodef{FC}{fusion center}
\acrodef{MAC}{multiple-access channel}
\acrodef{FL}{federated learning}
\acrodef{ED}{edge device}
\acrodef{CS}{compressed sensing}
\acrodef{ES}{edge server}
\acrodef{DCN}{data center network}
\acrodef{RIS}{reconfigurable intelligent surfaces}
\acrodef{IMC}{in-memory computing}
\acrodef{FPGA}{field-programmable gate array}
\acrodef{SDR}{software-defined radio}
\acrodef{PS}{processing system}
\acrodef{SS}{soft synchronization}
\acrodef{IQ}{in-phase/quadrature}
\acrodef{IP}{intellectual property}
\acrodef{DMA}{direct-memory access}
\acrodef{RAM}{random access memory}
\acrodef{CC}{companion computer}
\acrodef{FEE}{function estimation error}
\acrodef{MSK}{minimum-shift keying}
\acrodef{TDMA}{time-domain multiple access}
\acrodef{PLNC}{physical-layer network coding}
\acrodef{UAV}{unmanned aerial vehicle}
\acrodef{LoRa}{Long-Range}
\acrodef{DC}{direct-current}
\acrodef{DAC}{digital-to-analog converter}
\acrodef{ADC}{anlog-to-digital converter}
\acrodef{CS}{complementary sequence}
\acrodef{GCP}{Golay complementary pair}
\acrodef{ANF}{algebraic normal form}
\acrodef{AACF}{aperiodic auto-correlation function}
\acrodef{RM}{Reed-Muller}

\acrodef{PUCCH}{physical uplink control channel}

\acrodef{OBO}{output-power back-off}
\acrodef{ACLR}{adjacent-channel-leakage ratio}

\acrodef{LDPC}{low-density parity check}

\acrodef{PDF}{probability density function}
\acrodef{CDF}{cummulative distribution function}

\acrodef{TBMA}{type-based multiple access}

\acrodef{MSFE}{mean-squared function error}
\acrodef{FEE}{function-estimation error}
\acrodef{CER}{computation error rate}
\acrodef{BCER}{block-computation error rate}
\acrodef{CFO}{carrier frequency offset}
\acrodef{TO}{time offset}
\acrodef{PO}{phase offset}
\acrodef{RSSI}{received signal strength  information}

\acrodef{STLC}{space-time line code}
\acrodef{CCI}{co-channel interference}
\acrodef{CSIT}[CSIT]{\ac{CSI} at the transmitter}
\acrodef{CSIR}[CSIR]{\ac{CSI} at the receiver}
\acrodef{MIMO}{multiple-input-multiple-output}
\acrodef{PC}{phase correction}
\acrodef{ZF}{zero-forcing}
\acrodef{ANOVA}{analysis of variance}

\acrodef{PCA}{principal component analysis}
\acrodef{TIG}{Technical Interest Group}

\acrodef{FSK}{frequency-shift keying}
\acrodef{PPM}{pulse-position modulation}
\acrodef{PAM}{pulse-amplitude modulation}

\acrodef{MRC}{maximum-ratio combining}
\acrodef{HP}{hard-coded participation}
\acrodef{HPA}{hard-coded participation with absentees}
\acrodef{SP}{soft-coded participation}
\acrodef{FSK-MV}{\ac{FSK}-based \ac{MV}}
\acrodef{RF}{radio-frequency}
\acrodef{MF}{matched filter}
\acrodef{PPM}{pulse-position modulation}
\acrodef{CSK}{chirp-shift keying}
\acrodef{PPM-MV}[PPM-MV]{\ac{PPM}-based \ac{MV}}
\acrodef{DFT-s-OFDM}{\ac{DFT}-spread \ac{OFDM}}
\acrodef{SC}{single-carrier}
\acrodef{SGD}{stochastic gradient descent}
\acrodef{signSGD}{sign stochastic gradient descent}

\acrodef{SL}{split learning}
\acrodef{SNR}{signal-to-noise ratio}
\acrodef{RMSE}{root-mean-squared error}
\acrodef{OFDM}{orthogonal frequency division multiplexing}
\acrodef{DFT}{discrete Fourier transform}
\acrodef{PSK}{phase-shift keying}
\acrodef{QAM}{quadrature amplitude modulation}
\acrodef{QPSK}{quadrature phase-shift keying}
\acrodef{PMEPR}{peak-to-mean envelope power ratio}
\acrodef{BER}{bit-error ratio}
\acrodef{SNR}{signal-to-noise ratio}
\acrodef{PSD}{power spectral density}
\acrodef{SE}{spectral efficiency}
\acrodef{CP}{cyclic prefix}
\acrodef{AWGN}{additive white Gaussian noise}
\acrodef{CFR}{channel frequency response}
\acrodef{CIR}{channel impulse response}
\acrodef{MMSE}{minimum mean-squared error}
\acrodef{LMMSE}{linear minimum mean-squared error}
\acrodef{BPSK}{binary phase shift keying}
\acrodef{BPSK}{quadrature phase shift keying}
\acrodef{BLER}{block-error rate}
\acrodef{ML}{maximum likelihood}
\acrodef{PHY}{physical layer}
\acrodef{PA}{power amplifier}
\acrodef{IDFT}{inverse DFT}
\acrodef{DoF}{degrees-of-freedom}
\acrodef{IoT}{Internet-of-Things}
\acrodef{FDE}{frequency-domain equalization}
\acrodef{RF}{radio-frequency}
\acrodef{IM}{index modulation}
\acrodef{MF}{matched filter}
\acrodef{PPM}{pulse-position modulation}

\acrodef{MSE}{mean-squared error}
\acrodef{MRT}{maximum-ratio transmission}
\acrodef{ERC}{equal-ratio combining}
\acrodef{BAA}{broadband analog aggregation}
\acrodef{OBDA}{one-bit broadband digital aggregation}
\acrodef{FEEL}{federated edge learning}
\acrodef{FL}{federated learning}
\acrodef{UL}{uplink}
\acrodef{DL}{downlink}
\acrodef{OAC}{over-the-air computation}
\acrodef{TCI}{truncated-channel inversion}
\acrodef{MV}{majority vote}
\acrodef{CNN}{convolution neural network}
\acrodef{ReLU}{rectified-linear unit}
\acrodef{CSI}{channel state information}
\acrodef{PAPR}{peak-to-average power ratio}
\acrodef{SC}{single-carrier}
\acrodef{iid}[IID]{independent and identically distributed}
\acrodef{RMS}{root-mean-square}
\acrodef{4G}{Fourth Generation}
\acrodef{5G}{Fifth Generation}
\acrodef{NR}{New Radio}
\acrodef{LTE}{Long-Term Evolution}
\acrodef{DFT-s-OFDM}{\ac{DFT}-spread \ac{OFDM}}
\acrodef{OFDMA}{orthogonal frequency division multiple access}
\acrodef{HARQ}{hybrid automatic repeat request}
\acrodef{D2D}{Device-to-Device}
\acrodef{NOMA}{non-orthogonal multiple access}
\acrodef{OMA}{orthogonal multiple access}

\acrodef{IMT}{International Mobile Telecommunications}
\acrodef{ITU}{International Telecommunication Union}

\acrodef{AAM}{adaptive absolute maximum}
\begin{document}

\title{
	{Wireless Federated $k$-Means Clustering with Non-coherent Over-the-Air Computation}
}

\author{
	\IEEEauthorblockN{Alphan \c{S}ahin} \IEEEauthorblockA{Electrical Engineering Department, University of South Carolina, Columbia, SC, USA\\
		Emails: asahin@mailbox.sc.edu}
} 

\maketitle

\begin{abstract}
In this study, we propose using an \ac{OAC} scheme for the federated $k$-means clustering algorithm  to reduce the per-round communication latency when it is implemented over a wireless network. The \ac{OAC} scheme relies on an encoder exploiting the representation of a number in a balanced number system and computes the sum of the updates for the federated $k$-means via signal superposition property of wireless multiple-access channels non-coherently to eliminate the need for precise phase and time synchronization. Also, a re-initialization method for ineffectively used centroids is proposed to improve the performance of the proposed method for heterogeneous data distribution. For a customer-location clustering scenario, we demonstrate the performance of the proposed algorithm and compare it with the standard $k$-means clustering. Our results show that the proposed approach performs similarly to the standard $k$-means while reducing communication latency.
\end{abstract}
\section{Introduction}
\acresetall

\Ac{OAC} is a physical layer concept that can benefit a wide variety of applications for function computation over a bandwidth-limited wireless channel  by reducing resource utilization to a one-time cost  that does not scale with the number of \acp{ED}  \cite{sahinSurvey2023}. It exploits the signal superposition property of wireless multiple-access channels to compute a set of special mathematical functions such as arithmetic mean and sum \cite{Nazer_2007,goldenbaum2013harnessing,goldenbaum2015nomographic}. With the increased attention to computation-oriented applications over wireless networks, \ac{OAC} has been utilized as a fundamental tool to improve communication latency.  For example,  in \cite{Sahin_2022MVjournal,Guangxu_2020,Guangxu_2021}, \ac{OAC} is used for aggregating gradients or model parameters of neural networks for supervised distributed training, such as \ac{FL} \cite{pmlr-v54-mcmahan17a},  over a wireless network to improve per-round communication latency. With the same motivation, in this work, we investigate \ac{OAC} for obtaining an unsupervised federated learning algorithm, i.e., the federated $k$-means algorithm, over wireless networks.

The $k$-means algorithm is a well-known algorithm that successively partitions a dataset to improve a metric that measures cluster formation. In the literature, it has been analyzed for various distributed settings. For instance, in \cite{Jagannathan_2005}, the authors introduce a privacy-preserving protocol, which relies on exchanging the centroids between two parties with vertically- or horizontally-partitioned data. The federated $k$-means algorithm is first explicitly mentioned in \cite{Kumar_2020}, where the authors apply it to a clustering task based on MNIST and EMNIST datasets. In \cite{Dennis_20221pmlr}, a one-shot federated clustering scheme is proposed. In this method, the \acp{ED} run the $k$-means locally and send the clustering results to the \ac{ES} for aggregation.  In \cite{Ghosh_2022tit}, one-shot federated clustering is extended to an iterative federated clustering algorithm.  For guaranteeing the privacy of federated $k$-means, in \cite{li2022secure}, it is proposed to use Lagrange encoding on local data and share
 the coded data samples across the \acp{ED} along with noise injection. In  \cite{Stallmann_2022mdpi}, a federated clustering framework that determines the number of global clusters and validates the clustering via Davies–Bouldin index. In \cite{Zhou2022}, the memory and communication efficiency of the federated $k$-means is proposed to be reduced by using the low-dimensional features of the local data samples. In \cite{Yang_2023}, it is proposed to initialize the centroids at the EDs for better centroid initialization. To the best of our knowledge, the federated $k$-means algorithm over a wireless network with \ac{OAC} is not investigated in the literature.

In this study, we propose to implement the federated $k$-means algorithm over wireless networks with a non-coherent \ac{OAC} based on balanced number systems \cite{sahin2023md}. The proposed approach reduces per-round communication latency by computing the sum of the local updates for clustering over the air while promoting data privacy via federation. To improve the performance of clustering while taking the data heterogeneity into account, we use a maximum adaptation approach for the OAC scheme and employ a simple-but-effective re-initialization method for the centroids that have small number of data samples. We compare the proposed algorithm with the case when the global dataset is available at a central server for various OAC configurations under different channel conditions.

\section{System Model}
\label{sec:system}
\subsection{Problem Statement}
\begin{figure}[t]
	\centering
	{\includegraphics[width =2.6in]{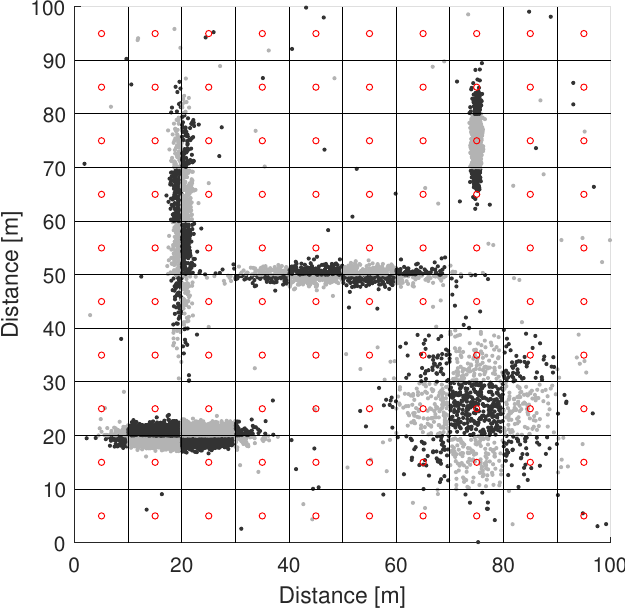}
	} 
	\caption{A clustering scenario. Each tile corresponds to a retail store in a mall, where each store has a local dataset containing their customers' locations (black and gray points). The stores are interested in where the customers are clustered in the mall without uploading their datasets to a central server.}
	\label{fig:scenario}
\end{figure}

Consider a scenario  where $\numberOfEdgeDevices$ \acp{ED} are connected to an \ac{ES} over a wireless network. Let $\dataset[\indexED]$ denote the dataset available at the $\indexED$th \ac{ED}, where a data sample $\dataSample[]$ in $\dataset[\indexED]$ is an ${\lengthOfDataSample}$-dimensional real vector, $\forall\indexED$. Suppose that the \acp{ED} are not willing to share their datasets with the \ac{ES} due to privacy considerations. Under this constraint, the objective of each ED is to learn where data samples are clustered in  the global dataset, i.e., $\completeData=\dataset[1]\cup\dataset[2]\cup\cdots\cup\dataset[\numberOfEdgeDevices]$, for further inference. For instance, consider the rectangular tessellation given in \figurename~\ref{fig:scenario} with $100$ tiles, where each tile corresponds to a retail store in a mall. Each retail store has a dataset that contains the precise $x$- and $y$-coordinates of their customers' locations, i.e., the points that reside within the corresponding tile. For this scenario, to assess the customers' preferences, each store is interested in where the points (e.g., customers' locations) are clustered in the entire mall without uploading their datasets to a central server. For this example, 
the EDs may be local radios at the retail stores, connected to a base station, i.e., an ES, located at the center of the mall. As can be seen from \figurename~\ref{fig:scenario}, the data distributions and the cardinalities of  datasets at the EDs can widely vary since  each dataset contains only the customers' positions within the store. 

The aforementioned scenario can be expressed as an optimization problem seeking for the  centroids of $\numberOfCentroids$ disjoint clusters $\partition[1],\mydots,\partition[\numberOfCentroids]$ that
partition the set $\completeData$ to minimize a loss function given by
\begin{align}
\lossFunction[{\partition[1],\mydots,\partition[\numberOfCentroids]}]=\sum_{\indexCentroid=1}^{\numberOfCentroids}\sum_{\dataSample[]\in\partition[\indexCentroid]}\norm*{\dataSample[]-\meanPartition[\indexCentroid]}_2^2~,
	\label{eq:originalProblem}
\end{align}
where $\meanPartition[\indexCentroid] = \frac{1}{\left|\partition[\numberOfCentroids]\right|} \sum_{\dataSample[]\in\partition[\indexCentroid]}\dataSample[]$ is the centroid of $\indexCentroid$th cluster. The minimization of \eqref{eq:originalProblem} is NP-hard \cite{Garey_1982}. Hence, we consider an approximate solution via the $k$-means clustering algorithm.

 The standard $k$-means algorithm aims to solve \eqref{eq:originalProblem} iteratively. For a given a set of centroids $\centroid[1][\indexCommunicationRound],\mydots,\centroid[\numberOfCentroids][\indexCommunicationRound]$, it calculates the $\indexCentroid$th cluster based on Euclidean distance as
$
\partitionIterations[\indexCentroid][\indexCommunicationRound]=\{\dataSample[]|\norm{\dataSample[]-\centroid[\indexCentroid][\indexCommunicationRound]}_2^2\le\norm{\dataSample[]-\centroid[\indexCentroidAnother][\indexCommunicationRound]}_2^2,\forall\indexCentroidAnother\}
$. Subsequently, the $\indexCentroid$th centroid is updated as
\begin{align}
	\centroid[\indexCentroid][\indexCommunicationRound+1]&=(1-\learningRate)\centroid[\indexCentroid][\indexCommunicationRound]+\learningRate\frac{1}{|\partitionIterations[\indexCentroid][\indexCommunicationRound]|}\sum_{\dataSample[]\in\partitionIterations[\indexCentroid][\indexCommunicationRound]}\dataSample[]~,
	\label{eq:updateOriginal}
\end{align}
for $|\partitionIterations[\indexCentroid][\indexCommunicationRound]|>0$, where $\learningRate$ is the learning rate and equal to $1$ for the standard $k$-means algorithm.

For our scenario, the partition $\partitionIterations[\indexCentroid][\indexCommunicationRound]$, $\forall\indexCentroid$, cannot be formed since the global dataset is not available at the ES. With {\em federated} $k$-means, clustering can be achieved without the global dataset  as follows: The ES distributes $\centroid[1][\indexCommunicationRound],\mydots,\centroid[\numberOfCentroids][\indexCommunicationRound]$ to the EDs for the $\indexCommunicationRound$th iteration. Given the centroids, each ED computes the local clusters based on its local dataset as
\begin{align}
	\partitionIterations[\indexED,\indexCentroid][\indexCommunicationRound]=\{\dataSample[]|\norm{\dataSample[]-\centroid[\indexCentroid][\indexCommunicationRound]}_2^2\le\norm{\dataSample[]-\centroid[\indexCentroidAnother][\indexCommunicationRound]}_2^2,\forall\indexCentroidAnother,\dataSample[]\in\dataset[\indexED]\}~.
	\label{eq:localPartitions}
\end{align}
The update step in \eqref{eq:updateOriginal} can then be re-expressed as
\begin{align}
	\centroid[\indexCentroid][\indexCommunicationRound+1]&=(1-\learningRate)\centroid[\indexCentroid][\indexCommunicationRound]+\learningRate\frac{1}{\left|\partitionIterations[\indexCentroid][\indexCommunicationRound]\right|}\sum_{\indexED=1}^{\numberOfEdgeDevices}\sum_{\dataSample[]\in\partitionIterations[\indexED,\indexCentroid][\indexCommunicationRound]}\dataSample[]\label{eq:fedSumData}\\
	&=\centroid[\indexCentroid][\indexCommunicationRound]+\learningRate\frac{1}{\left|\partitionIterations[\indexCentroid][\indexCommunicationRound]\right|}\sum_{\indexED=1}^{\numberOfEdgeDevices}
	\localUpdateVector[\indexED,\indexCentroid][\indexCommunicationRound]~,\label{eq:fedUpdateData}
\end{align}
for $|\partitionIterations[\indexCentroid][\indexCommunicationRound]|=\sum_{\indexED=1}^{\numberOfEdgeDevices}|\partitionIterations[\indexED,\indexCentroid][\indexCommunicationRound]|>0$ and $\localUpdateVector[\indexED,\indexCentroid][\indexCommunicationRound]$ is defined by
\begin{align}
	\localUpdateVector[\indexED,\indexCentroid][\indexCommunicationRound] \triangleq \sum_{\dataSample[]\in\partitionIterations[\indexED,\indexCentroid][\indexCommunicationRound]}\dataSample[]-\centroid[\indexCentroid][\indexCommunicationRound]~.
	\label{eq:updateSum}
\end{align}
Thus, with the federated $k$-means algorithm,  the $\indexED$th ED shares either the sum of data samples within the cluster or the total change with the ES, as can be seen in \eqref{eq:fedSumData} and \eqref{eq:fedUpdateData}, respectively. Since the data samples are not shared with the federated $k$-means algorithm, the privacy is improved at the expense of per-round communication latency (or resource utilization) that grows linearly with the number of \acp{ED} due to the communication between the EDs and the ES.
In this work, we address the latency issue of the federated $k$-means over wireless networks with \ac{OAC}.

\subsection{Signal model and wireless channel}
We assume that each \ac{ED} and the \ac{ES}  are equipped with a single antenna and the large-scale impact of the wireless channel is compensated with a state-of-the-art power control mechanism \cite{10.5555/3294673}. For the signal model, we assume that the \acp{ED} access the wireless channel  on the same time-frequency resources {\em simultaneously} with \ac{OFDM} symbols.  
Assuming that the \ac{CP} duration is larger than the sum of the maximum time-synchronization error and the maximum-excess delay of the channel,  the  received symbol on the $\indexSubcarrier$th resource (e.g., an \ac{OFDM} subcarrier) can be expressed as 
\begin{align}
	\receivedSymbolAtSubcarrier[{\indexSubcarrier}] = \sum_{\indexED=1}^{\numberOfEdgeDevices} \channelAtSubcarrier[\indexED,\indexSubcarrier]\transmittedSymbolAtSubcarrier[\indexED,\indexSubcarrier]+\noiseAtSubcarrier[\indexSubcarrier]~,
	\label{eq:symbolOnSubcarrier}\end{align}
where  $\channelAtSubcarrier[\indexED,\indexSubcarrier]\sim\complexGaussian[0][1]$ is the channel coefficient between the \ac{ES} and the $\indexED$th \ac{ED}, $\transmittedSymbolAtSubcarrier[\indexED,\indexSubcarrier]\in\complexNumbers$ is the  transmitted symbol from the $\indexED$th \ac{ED}, and ${\noiseAtSubcarrier[\indexSubcarrier]}\sim\complexGaussian[0][{\noiseVariance}]$ is zero-mean symmetric \ac{AWGN} with the variance  $\noiseVariance$.   $\SNR=1/\noiseVariance$ denotes the \ac{SNR} of an \ac{ED} at the \ac{ES} receiver. 
 
\def\localGradientElement[#1][#2]{{v}_{#1}^{(#2)}} 
\def\localGradientElementQuantized[#1][#2]{{\bar{v}}_{#1}^{(#2)}}
\section{Federated $k$-Means with Non-coherent OAC}
\label{sec:scheme}
In this section, we discuss how we address the communication bottleneck of wireless federated $k$-means by computing the sum in \eqref{eq:fedUpdateData} with a non-coherent \ac{OAC} scheme without using the \ac{CSI}, i.e., $\channelAtSubcarrier[\indexED,\indexSubcarrier]$, $\forall\indexED$, $\forall\indexSubcarrier$, at the \acp{ED} and \ac{ES}. To this end, we consider the OAC scheme that exploits balanced number systems \cite{sahin2023md}.
\subsection{Edge Device - Transmitter}
 Let $\localGradientElement[\indexED,\indexGradient][\indexCommunicationRound]$ be the $(\indexGradient+1)$th element of $\vectorization{[\localUpdateVector[\indexED,1][\indexCommunicationRound], \mydots, \localUpdateVector[\indexED,\numberOfCentroids][\indexCommunicationRound]]}\in\realNumbers^{\lengthOfDataSample\numberOfCentroids}$ for $\indexGradient\in\{0,1,\mydots,\lengthOfDataSample\numberOfCentroids-1\}$, where $\vectorization{\cdot}$ is the vectorization operation. The $\indexED$th \ac{ED} encodes $\localGradientElement[\indexED,\indexGradient][\indexCommunicationRound]$ into a sequence of length $\numberOfDigits$ as
\begin{align}
	\representationInBase[\base][{\digit[\indexED,\indexGradient][\numberOfDigits-1],\mydots,\digit[\indexED,\indexGradient][\indexDigit],\mydots,\digit[\indexED,\indexGradient][0]}]=\encoder(\localGradientElement[\indexED,\indexGradient][\indexCommunicationRound])~,
	\label{eq:encodedSequence}
\end{align}
for $\digit[\indexED][\indexDigit]\in	\symbolSet[\base]
\triangleq\{\symbol[{\indexSymbol}]|\symbol[{\indexSymbol}]=\indexSymbol-(\base-1)/2, \indexSymbol\in\{0,1,\mydots,\base-1\}\}$, $\forall\indexDigit$, where $\base$ is an odd positive integer (i.e., base) and $\encoder$ is a function that maps $\localGradientElement[\indexED,\indexGradient][\indexCommunicationRound]$ to a sequence  of  $\numberOfDigits$ numerals in a balanced number system with base $\base$. The numerals are obtained via  $\encoder(\localGradientElement[\indexED,\indexGradient][\indexCommunicationRound])$  as follows \cite{sahin2023md}: 
\begin{enumerate}
	\item  $\localGradientElement[\indexED,\indexGradient][\indexCommunicationRound]$ is clamped
	as $\aRealValueClamped=\max(-\valueMaximum, \min(\localGradientElement[\indexED,\indexGradient][\indexCommunicationRound],\valueMaximum))$ to ensure $\aRealValueClamped\in[-\valueMaximum,\valueMaximum]$ for a given $\valueMaximum>0$.
	\item $\aRealValueClamped$ is re-scaled as $\frac{\normalizationDigit}{\valueMaximum}\aRealValueClamped+\normalizationDigit+\frac{1}{2}$. 
	\item The scaled value is mapped to an integer between $0$ and $2\normalizationDigit$ with a floor operation  and the corresponding integer is expanded as $
		\floor*{\frac{\normalizationDigit}{\valueMaximum}\aRealValueClamped+\normalizationDigit+\frac{1}{2}}=\sum_{\indexDigit=0}^{\numberOfDigits-1}\digitStandart[\indexDigit]\base^\indexDigit$,	for $\digitStandart[\indexED,\indexDigit]\in\integers_\base$ and $\normalizationDigit\triangleq{(\base^{\numberOfDigits}-1)}/{2}$.
	\item $\digit[\indexED,\indexGradient][\indexDigit]$ is calculated as $
	\digit[\indexED,\indexGradient][\indexDigit]=\digitStandart[\indexDigit]-(\base-1)/2$, $\forall\indexDigit$.
\end{enumerate}
It is worth noting that 
the quantized $\localGradientElement[\indexED,\indexGradient][\indexCommunicationRound]$ can be obtained  as
\begin{align}
	\localGradientElementQuantized[\indexED,\indexGradient][\indexCommunicationRound]=\decoder	\representationInBase[\base][{\digit[\indexED,\indexGradient][\numberOfDigits-1],\mydots,\digit[\indexED,\indexGradient][0]}]\triangleq\frac{\valueMaximum}{\normalizationDigit}\sum_{\indexDigit=0}^{\numberOfDigits-1}\digit[\indexED,\indexGradient][\indexDigit]\base^{\indexDigit}~.\label{eq:decoderDef}
\end{align}
We refer the reader to \cite{sahin2023md} for several numerical examples with $\encoder$ and $\decoder$.

Without loss of generality,  in this study, we use a resource mapping rule given by $\indexSubcarrier=\base\numberOfDigits\indexGradient+\base\indexDigit+\indexSymbol$ for a given triplet $(\indexGradient,\indexDigit,\indexSymbol)$. Based on the numerals obtained in \eqref{eq:encodedSequence}, we compute the transmitted symbol $\transmittedSymbolAtSubcarrier[\indexED,\indexSubcarrier]$ in \eqref{eq:symbolOnSubcarrier}  as
\begin{align}
	\transmittedSymbolAtSubcarrier[\indexED,\indexSubcarrier]=
	\sqrt{\symbolEnergy} \randomSymbolAtSubcarrier[\indexED,\indexSubcarrier]\times\indicatorFunction[{\digit[\indexED,\indexGradient][\indexDigit]=\symbol[\indexSymbol]}]~,
	\label{eq:symbolPlus}
\end{align}
for $\symbol[\indexSymbol]\in\symbolSet[\base]$, where $\symbolEnergy\triangleq\sqrt{\base}$ is the symbol energy, $\randomSymbolAtSubcarrier[\indexED,\indexSubcarrier]$ is a random \ac{QPSK} symbol to improve the \ac{PMEPR} of the corresponding \ac{OFDM} waveform, and the function $\indicatorFunction[\cdot]$ results in $1$ if its argument holds, otherwise, it is $0$. Thus, with \eqref{eq:symbolPlus}, $\base$ complex-valued resources are dedicated to each numeral, and one of them is  activated  based on its value.

Since all EDs access the spectrum simultaneously  for OAC, the number of complex-valued resources consumed for each communication round can be calculated as $\lengthOfDataSample\numberOfCentroids\base\numberOfDigits$ and  not scaled with the number of EDs. Also, as the EDs do not use \ac{CSI}, not only the channel estimation overhead but also the need for phase and precise time synchronizations are eliminated with the aforementioned OAC scheme.  Note that, without OAC, the number of resources required may be roughly calculated as $\lengthOfDataSample\numberOfCentroids\numberOfEdgeDevices\spectralEfficiency\compressionRatio\numberOfBits$, where  $\spectralEfficiency$ is the spectral efficiency in bits/s/Hz, $\compressionRatio$ is the compression ratio, $\numberOfBits$ is the number of bits for representing $\localGradientElement[\indexED,\indexGradient][\indexCommunicationRound]$.

\begin{algorithm}[t]
	\caption{Wireless federated $k$-means with OAC}\label{alg:fedkmeansOAC}
	\footnotesize
	\SetKwInput{KwInput}{Input}                
	\SetKwInput{KwOutput}{Output}              
	\SetKwComment{Comment}{/* }{}
	
	\DontPrintSemicolon
	
	\KwInput{$\centroid[1][0],\mydots,\centroid[\numberOfCentroids][0],\valueMaximum^{(0)},\minimumCardinality,\learningRate,\metricFactor,\varianceCentroid,\base,\numberOfDigits,\communicationRounds$}
	\KwOutput{$\centroid[1][\communicationRounds],\mydots,\centroid[\numberOfCentroids][\communicationRounds]$}
	
	\SetKwFunction{FMain}{}
	{
		\For{$\indexCommunicationRound=1:\communicationRounds$} {
			\Comment*[l]{Processing @ EDs}
			\For{$\indexED=1:\numberOfEdgeDevices$}{
				Compute $\partitionIterations[\indexED,\indexCentroid][\indexCommunicationRound]$ with \eqref{eq:localPartitions}, $\forall\indexCentroid$\;				
				Compute $\localUpdateVector[\indexED,\indexCentroid][\indexCommunicationRound]$  with \eqref{eq:updateSum}, $\forall\indexCentroid$\;
				Compute $\transmittedSymbolAtSubcarrier[\indexED,\indexSubcarrier]$ with \eqref{eq:symbolPlus}, $\forall\indexSubcarrier$\;
			}
			\Comment*[l]{Superposition in the uplink}
			The EDs transmit the \ac{OFDM} symbols simultaneously for OAC\;
			The EDs transmit $|\partitionIterations[\indexED,\indexCentroid][\indexCommunicationRound]|$, $\metric[\indexED][\indexCommunicationRound]$, $\forall\indexCentroid$\;
			\Comment*[l]{Processing @ ES}
			Compute $\meanGradientEleEstimate[\indexCommunicationRound][\indexGradient]$ with \eqref{eq:finalValEst}, $\forall$ \;
			Update 	$\valueMaximum^{(\indexCommunicationRound+1)}$ with \eqref{eq:aam}\;			
			Update  $\centroid[\indexCentroid'][\indexCommunicationRound+1]$ with \eqref{eq:fedUpdateData}, $\forall\indexCentroid'\in\indicesNullSet^{\rm c}$\;  
			Update  $\centroid[\indexCentroid''][\indexCommunicationRound+1]$ with \eqref{eq:randomize}, $\forall\indexCentroid''\in\indicesNullSet$\; 	
			\Comment*[l]{Broadcast in the downlink}
			The ES broadcasts $\valueMaximum^{(\indexCommunicationRound+1)},\centroid[\indexCentroid][\indexCommunicationRound+1],\forall\indexCentroid$
		}
	}
\end{algorithm}
\subsection{Edge Server - Receiver}
At the \ac{ES}, we exploit the fact that
the $(\indexGradient+1)$th element of $\vectorization{\sum_{\indexED=1}^{\numberOfEdgeDevices}
\localUpdateVector[\indexED,1][\indexCommunicationRound], \mydots, \sum_{\indexED=1}^{\numberOfEdgeDevices}
\localUpdateVector[\indexED,\numberOfCentroids][\indexCommunicationRound]}$, denoted by $\meanGradientEle[\indexCommunicationRound][\indexGradient]$,  can be obtained approximately by using \eqref{eq:decoderDef} as
\begin{align}
	\meanGradientEle[\indexCommunicationRound][\indexGradient]=\sum_{\indexED=1}^{\numberOfEdgeDevices} \localGradientElement[\indexED,\indexGradient][\indexCommunicationRound]\approxeq\sum_{\indexED=1}^{\numberOfEdgeDevices} \localGradientElementQuantized[\indexED,\indexGradient][\indexCommunicationRound]&=\frac{\valueMaximum}{\normalizationDigit}\sum_{\indexDigit=0}^{\numberOfDigits-1} \sum_{\indexED=1}^{\numberOfEdgeDevices}\digit[\indexED,\indexGradient][\indexDigit]\base^{\indexDigit}\nonumber\\&=\decoder\representationInBase[\base][{\digitAveraged[\indexGradient][\numberOfDigits-1],\mydots,\digitAveraged[\indexGradient][0]}]~,\nonumber
\end{align}
for $
\digitAveraged[\indexGradient][\indexDigit]\triangleq\sum_{\indexED=1}^{\numberOfEdgeDevices}\digit[\indexED,\indexGradient][\indexDigit]= \sum_{\indexSymbol=0}^{\base-1} \symbol[\indexSymbol]\numberOFEDsForOptionGeneral[\indexSymbol]
$, where $\numberOFEDsForOptionGeneral[\indexSymbol]$ denotes the number of \acp{ED} given that the $\indexDigit$th numeral in \eqref{eq:encodedSequence} is $\symbol[\indexSymbol]$. Hence, we need to estimate $\numberOFEDsForOptionGeneral[\indexSymbol]$, $\forall\indexDigit$, $\forall\indexSymbol$, to obtain an estimate of $	\meanGradientEle[\indexCommunicationRound][\indexGradient]$. In \cite{sahin2023md},  it is shown that the norm of $\receivedSymbolAtSubcarrier[\indexSubcarrier]$ can be used as
$\numberOFEDsForOptionGeneralDetector[\indexSymbol]={(\norm{\receivedSymbolAtSubcarrier[\indexSubcarrier]}_2^2-\noiseVariance)}/\symbolEnergy$,
where $\indexGradient$, $\indexDigit$, and $\indexSymbol$ can be obtained as $\indexGradient=\floor{\indexSubcarrier/(\numberOfDigits\base)}$,  $\indexDigit=\floor{\indexSubcarrier/\base}\mod\numberOfDigits$, and $\indexSymbol=\indexSubcarrier\mod\base$, respectively, based on the resource mapping rule at the transmitters. Finally,  $\digitAveraged[\indexGradient][\indexDigit]$ and $\meanGradientEle[\indexCommunicationRound][\indexGradient]$  can be estimated as
$
	\digitAveragedEst[\indexGradient][\indexDigit]=
	\sum_{\indexSymbol=0}^{\base-1} \symbol[\indexSymbol]\numberOFEDsForOptionGeneralDetector[\indexSymbol]
$ and
\begin{align}
	\meanGradientEleEstimate[\indexCommunicationRound][\indexGradient]=\decoder\representationInBase[\base][{\digitAveragedEst[\indexGradient][\numberOfDigits-1],\mydots,\digitAveragedEst[\indexGradient][1],\digitAveragedEst[\indexGradient][0]}]~,
	\label{eq:finalValEst}	
\end{align}
respectively. Subsequently, $\centroid[\indexCentroid][\indexCommunicationRound]$ is updated via \eqref{eq:fedUpdateData}.  

In this study, we assume that each ED reports the cardinality of the local partitions, i.e., $\{|\partitionIterations[\indexED,\indexCentroid][\indexCommunicationRound]|,\forall\indexCentroid\}$, to the ES for  $|\partitionIterations[\indexCentroid][\indexCommunicationRound]|$ calculation. It is worth noting that the sum for computing $|\partitionIterations[\indexCentroid][\indexCommunicationRound]|$  can be evaluated with OAC for further resource saving.

\subsection{Enhancements}
\label{subsec:enhancements}
The performance of the wireless federated $k$-means with the proposed OAC scheme can be improved further with several methods. To reduce the quantization error, we adopt a similar protocol discussed in \cite{sahin2023md}  to  set $\valueMaximum$ adaptively. With this strategy, $\valueMaximum$ is updated throughout the communication rounds as
\begin{align}
	\valueMaximum^{(\indexCommunicationRound+1)}=\metricFactor\times\max_{\indexED}\metric[\indexED][\indexCommunicationRound]~,
	\label{eq:aam}
\end{align}
where $\metric[\indexED][\indexCommunicationRound]=\max_{\indexGradient}{\localGradientElement[\indexED,\indexGradient][\indexCommunicationRound]}$ is a single parameter transmitted to the ES over an orthogonal channel from the $\indexED$th \ac{ED}. 

\begin{figure*}
	\centering
	\subfloat[AWGN, $\SNR=20$~dB.]{\includegraphics[width =\figuresizeThird]{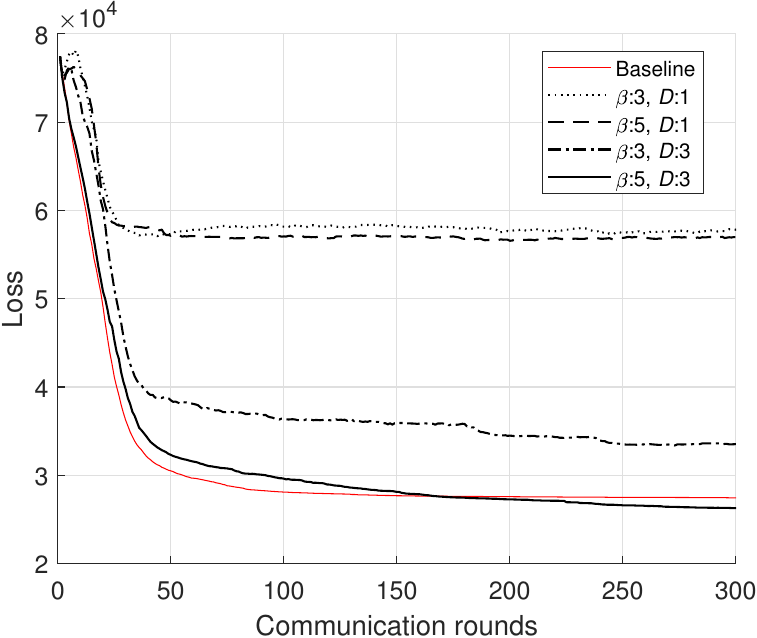}
		\label{subfig:awgn20_0c}}
	\subfloat[Flat fading, $\SNR=20$~dB.]{\includegraphics[width =\figuresizeThird]{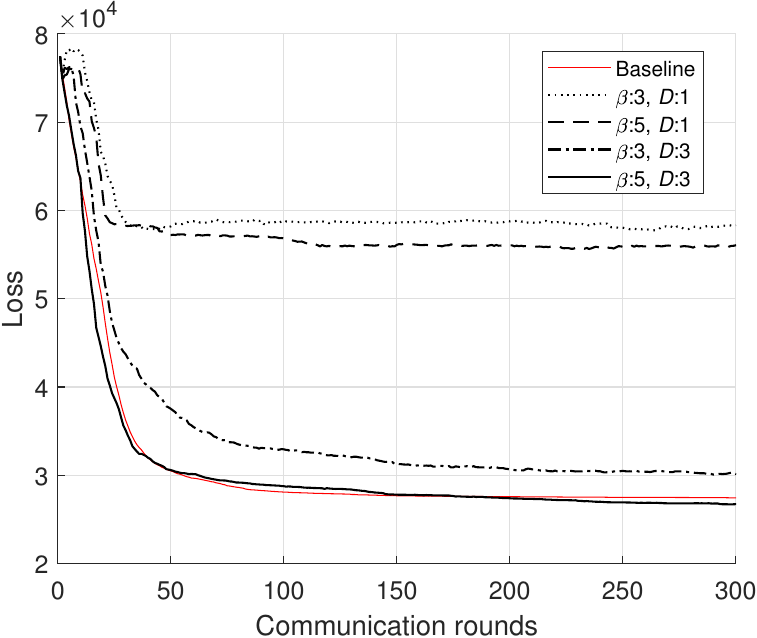}
		\label{subfig:flat20_0c}}
	\subfloat[Frequency-selective fading, $\SNR=20$~dB.]{\includegraphics[width =\figuresizeThird]{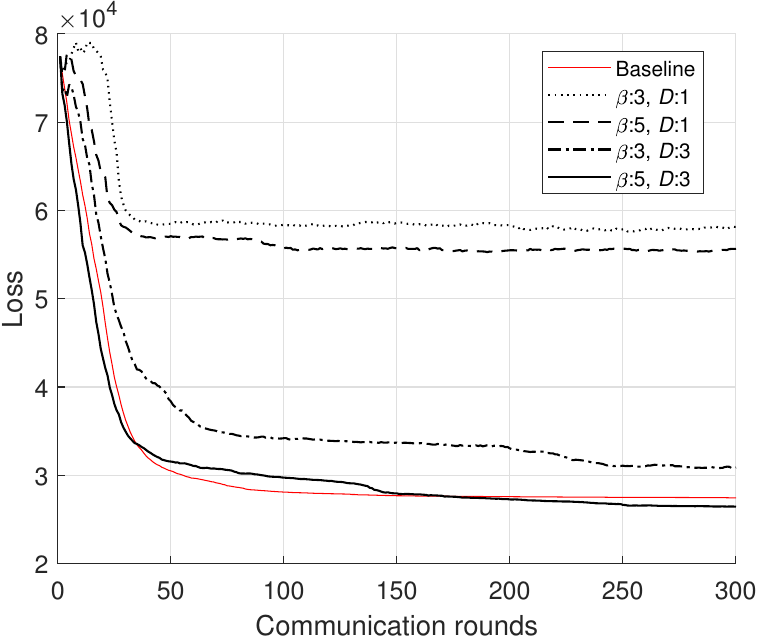}
		\label{subfig:sel20_0c}}		
	\\
	\subfloat[AWGN, $\SNR=10$~dB.]{\includegraphics[width =\figuresizeThird]{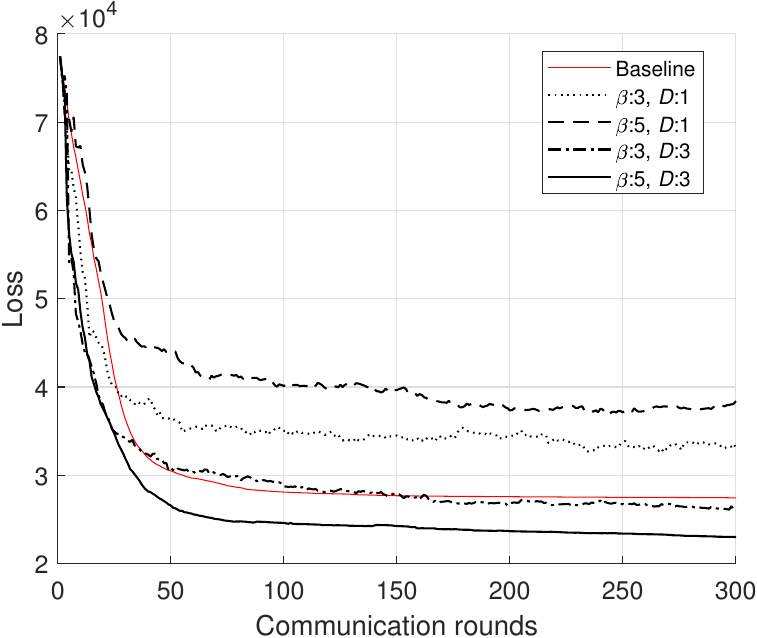}
		\label{subfig:awgn10_0c}}
	\subfloat[Flat fading, $\SNR=10$~dB.]{\includegraphics[width =\figuresizeThird]{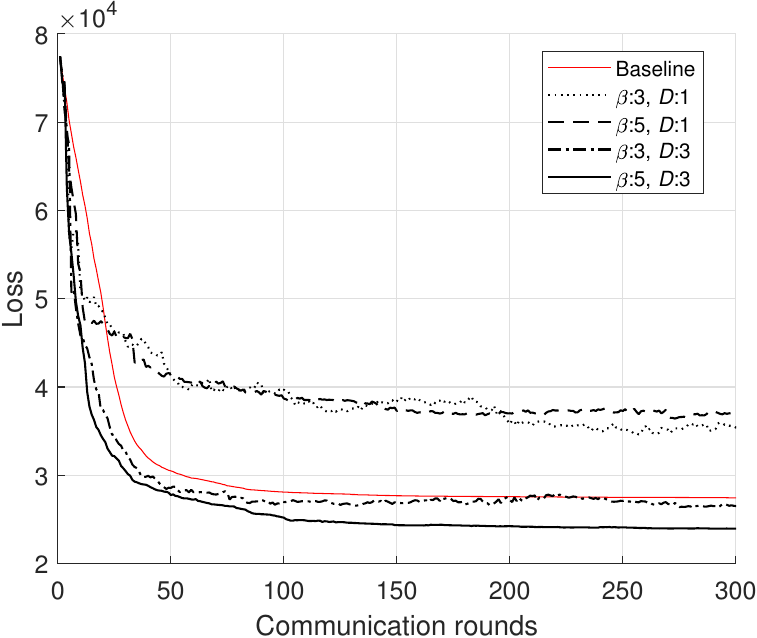}
		\label{subfig:flat10_0c}}
	\subfloat[Frequency-selective fading, $\SNR=10$~dB.]{\includegraphics[width =\figuresizeThird]{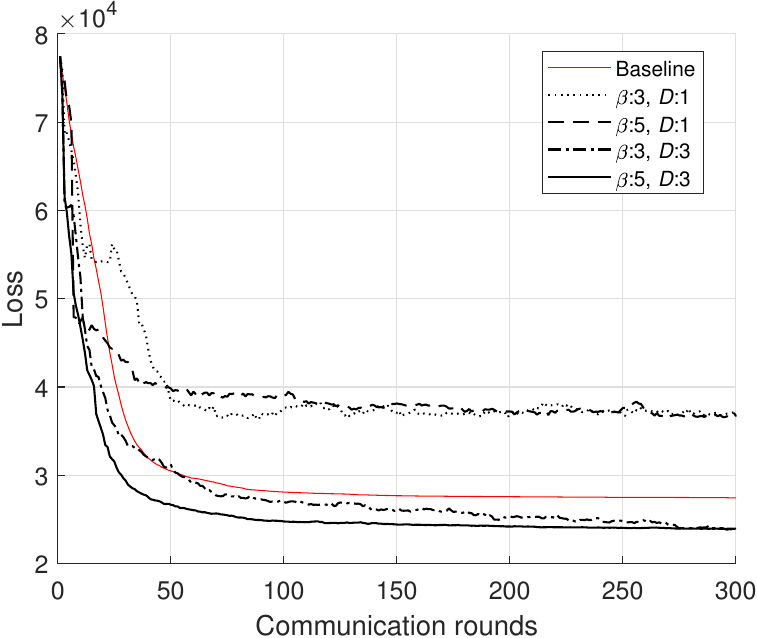}
		\label{subfig:sel10_0c}}
	\caption{The loss over the communication rounds for the wireless federated $k$-means with OAC ($\minimumCardinality=0$).}
	\label{fig:LossFedKmean0cardinality}
\end{figure*}
\begin{figure*}
	\centering
	\subfloat[AWGN, $\SNR=20$~dB.]{\includegraphics[width =\figuresizeThird]{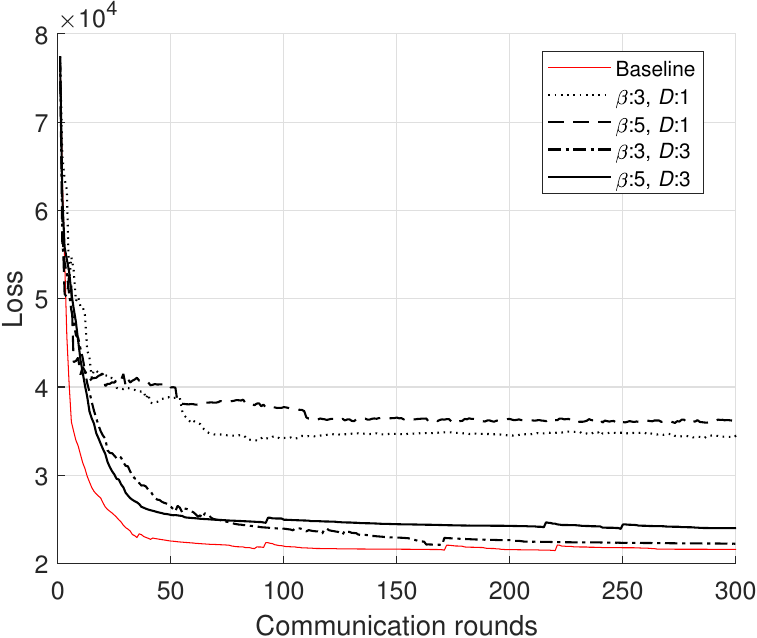}
		\label{subfig:awgn20_5c}}
	\subfloat[Flat fading, $\SNR=20$~dB.]{\includegraphics[width =\figuresizeThird]{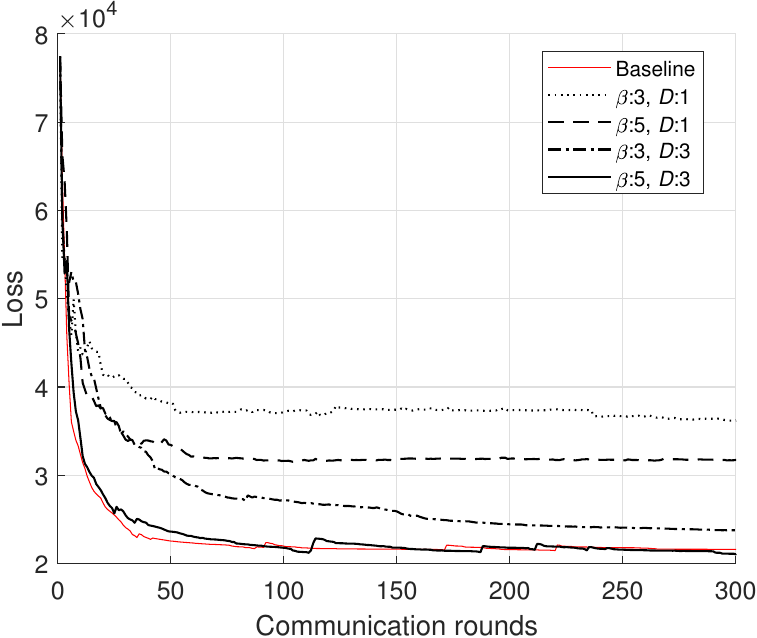}
		\label{subfig:flat20_5c}}
	\subfloat[Frequency-selective fading, $\SNR=20$~dB.]{\includegraphics[width =\figuresizeThird]{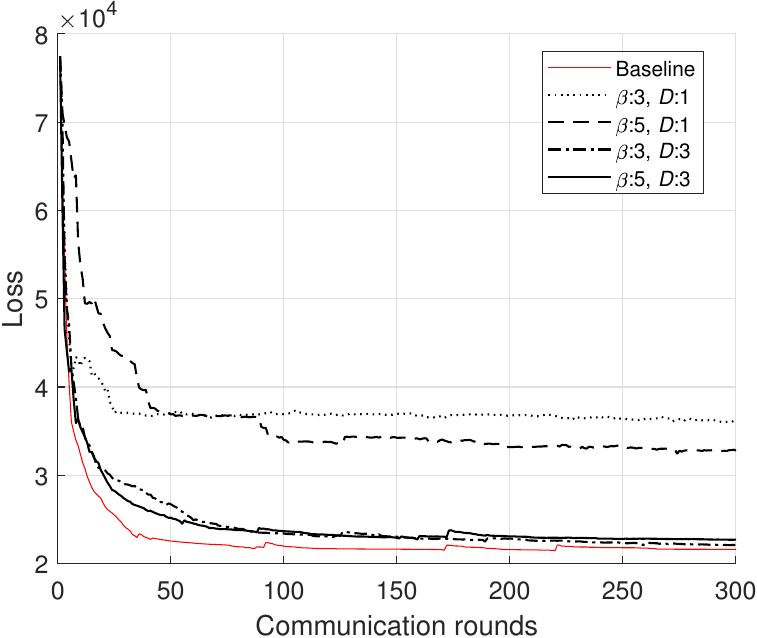}
		\label{subfig:sel20_5c}}		
	\\
	\subfloat[AWGN, $\SNR=10$~dB.]{\includegraphics[width =\figuresizeThird]{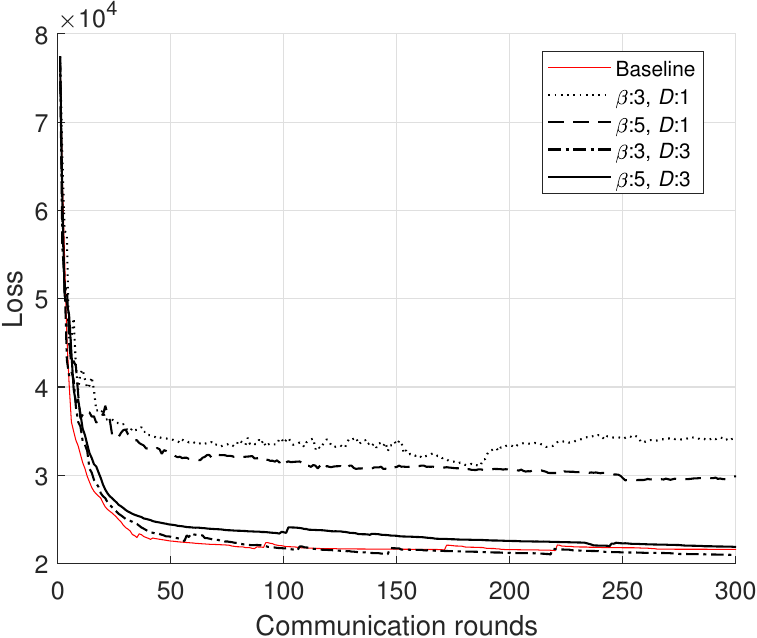}
		\label{subfig:awgn10_5c}}
	\subfloat[Flat fading, $\SNR=10$~dB.]{\includegraphics[width =\figuresizeThird]{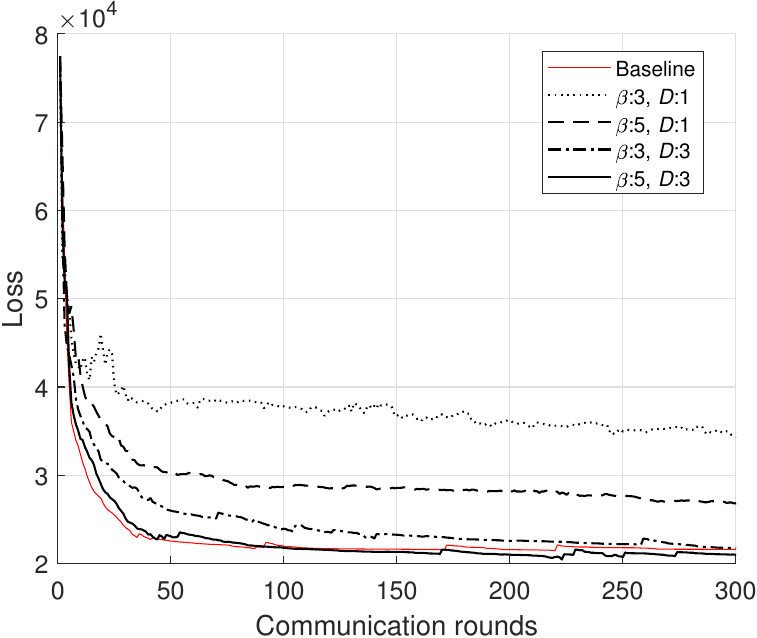}
		\label{subfig:flat10_5c}}
	\subfloat[Frequency-selective fading, $\SNR=10$~dB.]{\includegraphics[width =\figuresizeThird]{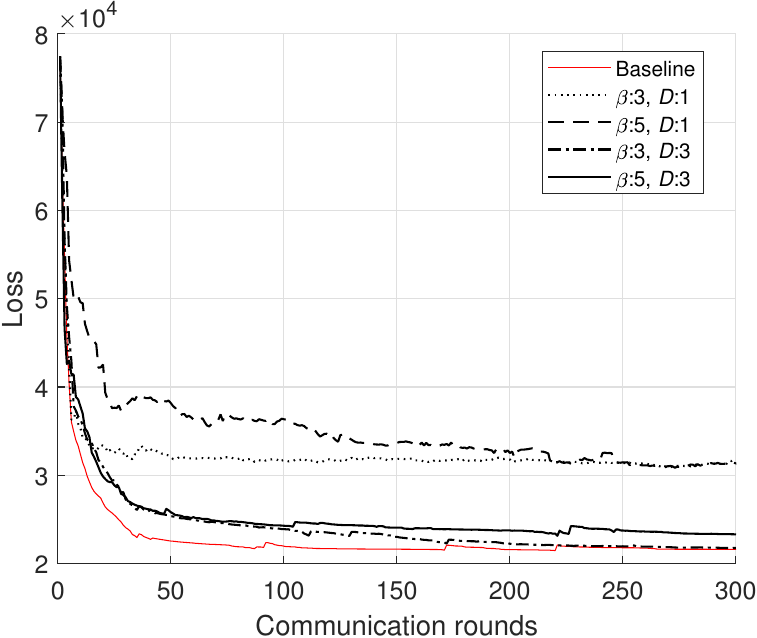}
		\label{subfig:sel10_5c}}
	\caption{The loss over the communication rounds for the wireless federated $k$-means with OAC ($\minimumCardinality=5$).}
	\label{fig:LossFedKmean5cardinality}
	\vspace{-1mm}
\end{figure*}
Since the ES does not know the global dataset, the cardinality of some of the partitions can be $0$. Hence, the corresponding centroids cannot be updated with \eqref{eq:fedUpdateData}. To address this issue, we introduce a generalized re-initialization step as
$
\centroid[\indexCentroid''][\indexCommunicationRound+1]=\centroid[\indexCentroid'][\indexCommunicationRound]+\centroidNoise[\indexCentroid''][\indexCommunicationRound]~,\label{eq:randomize}
$
for $\indexCentroid''\in\indicesNullSet\triangleq\{\indexCentroid|\partitionIterations[\indexCentroid][\indexCommunicationRound]|<\minimumCardinality, \minimumCardinality\ge0\}$, where $\indexCentroid'$ is chosen randomly from $\indicesNullSet^{\rm c}\triangleq\{1,\mydots,\numberOfCentroids\}/\indicesNullSet$ and $\centroidNoise[\indexCentroid''][\indexCommunicationRound]$ is a zero-mean random Gaussian vector with the variance of $\varianceCentroid$. With \eqref{eq:randomize},  the ES re-initializes a centroid where the cardinality of the corresponding partition is less than $\minimumCardinality$ by assigning it to a point nearby a centroid with $|\partitionIterations[\indexCentroid''][\indexCommunicationRound]|\ge\minimumCardinality$.

The corresponding algorithm with the aforementioned enhancements is given in Algorithm~\ref{alg:fedkmeansOAC}.

\section{Numerical Results}
\label{sec:numerical}
In this section, we analyze the performance of the wireless federated $k$-mean with the proposed OAC scheme for the scenario illustrated in \figurename\ref{fig:scenario}. We consider a $100~\text{m}$ $\times$ $100~\text{m}$ rectangular area for $\numberOfEdgeDevices=100$ EDs. We express the user's locations in a 2-D Cartesian coordinate system (i.e., $\lengthOfDataSample=2$) based on a mixture of Gaussian distributions (10000 points) and a uniform distribution (100 points). We choose the mixture weights, the mean values on the $x$- and $y$-axes, and the standard deviations on the $x$- and $y$-axes for the Gaussian mixture model as $(0.6, 20, 20, 5,1)$, $(0.1, 75, 25, 7, 7)$, $(0.1, 50, 50, 10, 1)$, $(0.1, 75, 75, 0.5, 4)$, and $(0.1, 20, 60, 1, 10)$. For the uniform distribution, we set the distribution boundaries  $0$ and $100$ meters for both $x$- and $y$-axes. For the algorithm, we consider $\numberOfCentroids=100$ clusters, where the initial values of the centroids are set to the center of the tiles, as shown in \figurename\ref{fig:scenario}. We choose $\valueMaximum^{(0)}=300$, $\minimumCardinality\in\{0,5\}$, $\learningRate=0.1$, $\metricFactor=1.2$, $\varianceCentroid=1$, $\base\in\{3,5\}$, and $\numberOfDigits=\{1,3\}$. We run the algorithm for $\communicationRounds=1000$ communication rounds. We generate the results for \ac{AWGN} channel (i.e., $\channelAtSubcarrier[\indexED,\indexSubcarrier]=1$, $\forall\indexED$, $\forall\indexSubcarrier$), flat fading channel (i.e., $\channelAtSubcarrier[\indexED,\indexSubcarrier]=\channelAtSubcarrier[\indexED,\indexSubcarrier']\sim\complexGaussian[0][1]$, $\forall\indexED$, $\forall\indexSubcarrier,\indexSubcarrier', \indexSubcarrier\neq\indexSubcarrier'$), and frequency-selective fading channel  (i.e., $\channelAtSubcarrier[\indexED,\indexSubcarrier]\sim\complexGaussian[0][1]$, $\forall\indexED$, $\forall\indexSubcarrier$) for $\SNR\in\{10,20\}$~dB. We regenerate the channel coefficients to model the time variation.  We compare our results with the standard $k$-means algorithm, denoted as the {\em baseline}, i.e., the scenario when $\dataset[]$ is available at the ES for clustering.

In \figurename~\ref{fig:LossFedKmean0cardinality}, we provide the loss in \eqref{eq:originalProblem} over the communication rounds for $\minimumCardinality=0$ and $\SNR=\{10, 20\}$~dB for different channel conditions. In this case, the re-initialization step discussed in Section~\ref{subsec:enhancements} is disabled. For $\numberOfDigits=1$, the OAC scheme introduces high quantization errors for both $\base=3$ and $\base=5$. Hence, for all channel conditions and SNR levels, their performances are worse than the cases for $\numberOfDigits=2$. The proposed scheme performs similarly to the baseline for $\numberOfDigits=2$ and $\base=5$. The performance of the proposed scheme is slightly better than that of the baseline due to the random noise in the communication channel, which allows the algorithm to find a better local optimum point. We observe a similar improvement when $\SNR$ is reduced to $20$~dB from $10$~dB.  In \figurename~\ref{fig:LossFedKmean5cardinality}, we analyze the same scenario in \figurename~\ref{fig:LossFedKmean0cardinality} for $\minimumCardinality=5$. In this case, the partitions have at least 5 data samples. Since the centroids are utilized more effectively, the loss is reduced further as compared to the ones in \figurename~\ref{fig:LossFedKmean0cardinality}. The simulation results vary marginally for different channel conditions, indicating that the wireless federated $k$-means with the OAC and the standard $k$-means can perform similarly when the quantization error is reduced by increasing $\numberOfDigits$ or $\base$. Also, with the proposed scheme, $\lengthOfDataSample\numberOfCentroids\base\numberOfDigits=2000$ complex-valued resources need to be utilized for computing the centroid updates. On the other hand, the same computation without OAC requires  $\lengthOfDataSample\numberOfCentroids\numberOfEdgeDevices\spectralEfficiency\compressionRatio\numberOfBits=32000$ for $\spectralEfficiency$~bits/s/Hz, $\compressionRatio=1/5$, and $\numberOfBits=8$.

\begin{figure}[t]
	\centering
	\subfloat[{Baseline.}]{\includegraphics[width =\figuresize/2]{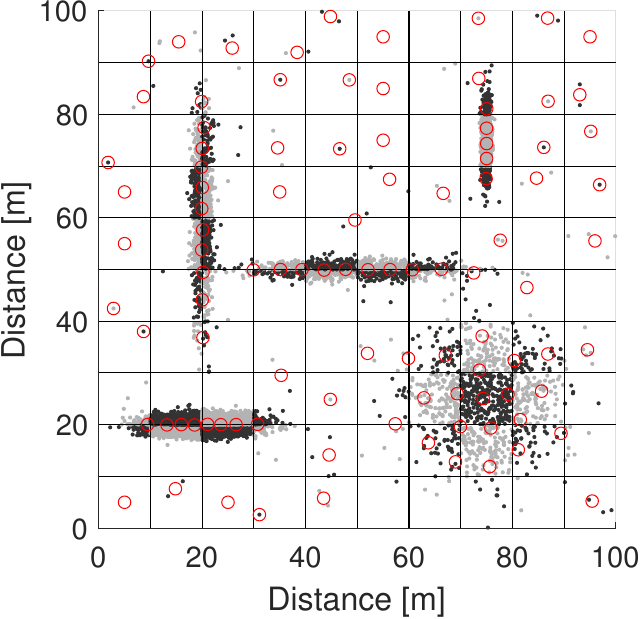}
		\label{subfig:baseline0c}}
	\subfloat[AWGN, {$\base=5$, $\numberOfDigits=3$.}]{\includegraphics[width =\figuresize/2]{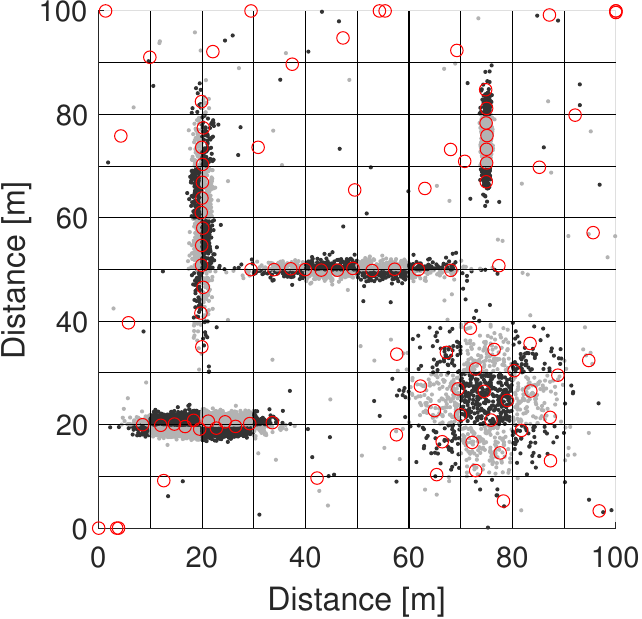}
		\label{subfig:awgn0c}}\\		
	\subfloat[{Flat fading, $\base=5$, $\numberOfDigits=3$.}]{\includegraphics[width =\figuresize/2]{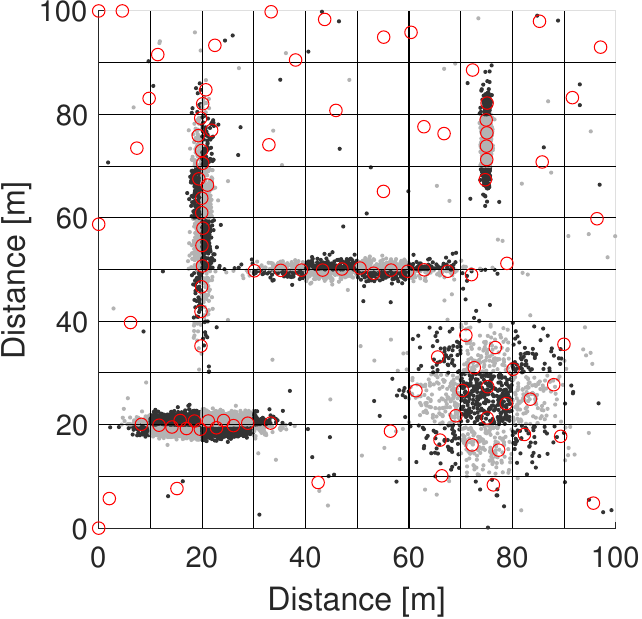}
		\label{subfig:flat0c}}
	\subfloat[{Freq.-sel. channel, $\base=5$, $\numberOfDigits=3$.}]{\includegraphics[width =\figuresize/2]{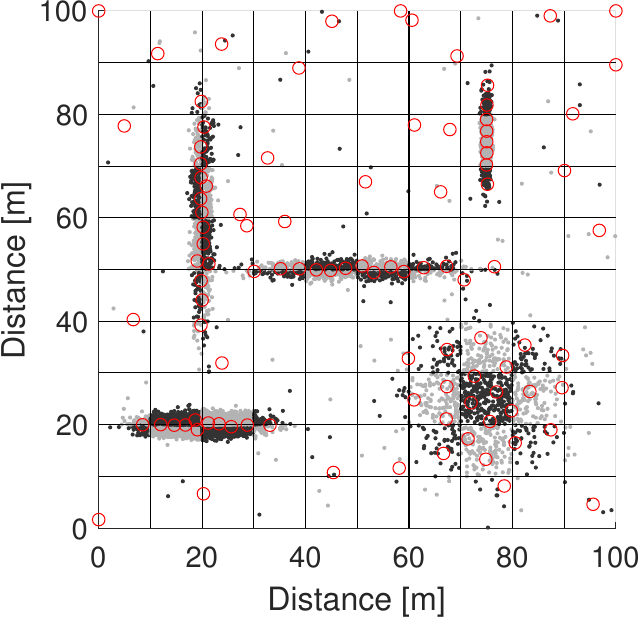}
		\label{subfig:selective0c}}		
	\caption{The final centroids for the wireless  federated $k$-means with OAC ($\SNR=10$~dB, $\minimumCardinality=0$).}
	\label{fig:depFedKmean0Card}
\end{figure}

\begin{figure}[t]
	\centering
	\subfloat[{Baseline.}]{\includegraphics[width =\figuresize/2]{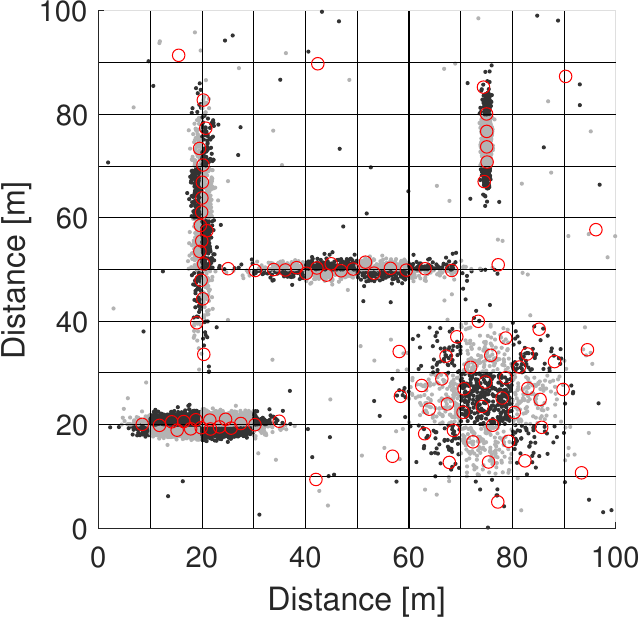}
		\label{subfig:baseline5c}}
	\subfloat[AWGN, {$\base=5$, $\numberOfDigits=3$.}]{\includegraphics[width =\figuresize/2]{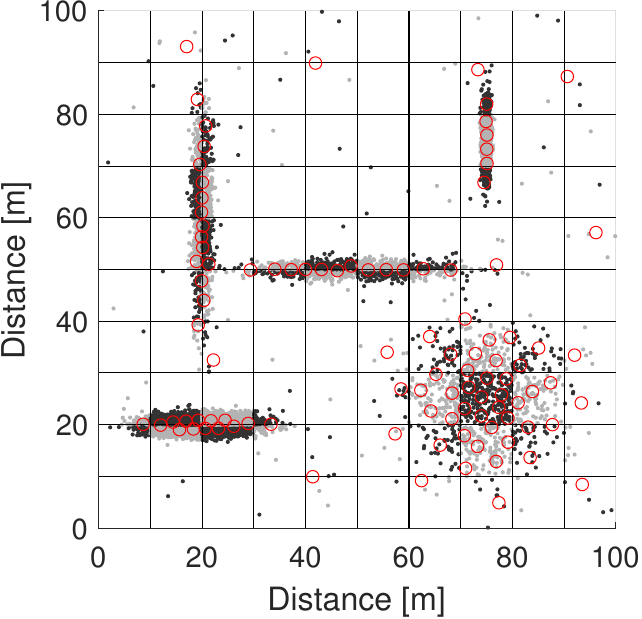}
		\label{subfig:awgn5c}}\\		
	\subfloat[{Flat fading, $\base=5$, $\numberOfDigits=3$.}]{\includegraphics[width =\figuresize/2]{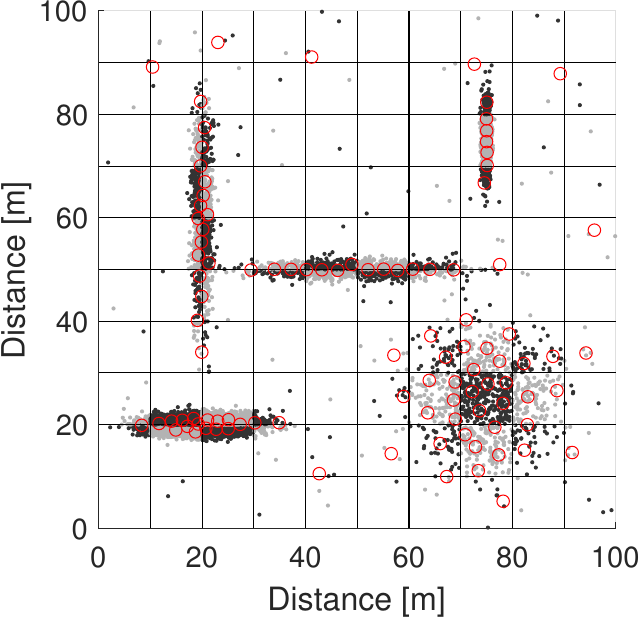}
		\label{subfig:flat5c}}
	\subfloat[{Freq.-sel. channel, $\base=5$, $\numberOfDigits=3$.}]{\includegraphics[width =\figuresize/2]{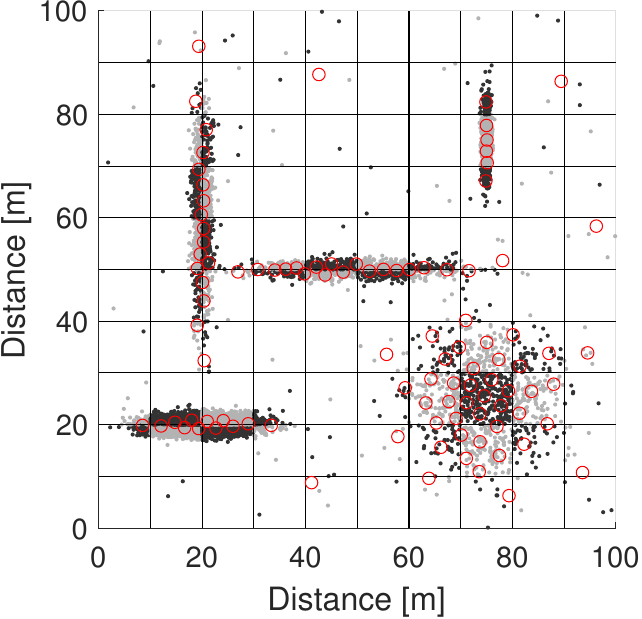}
		\label{subfig:selective5c}}		
	\caption{The final  centroids for the wireless federated $k$-means  with OAC ($\SNR=10$~dB, $\minimumCardinality=5$).}
	\label{fig:depFedKmean5Card}
\end{figure}

In \figurename~\ref{fig:depFedKmean0Card} and \figurename~\ref{fig:depFedKmean5Card}, we provide the  locations of the centroids after $\communicationRounds=1000$ communication rounds for $\minimumCardinality=0$ and $\minimumCardinality=5$. As can be seen, the centroid locations are similar to each other in different channel conditions. We observe that some of the centroids do not change their locations as the local datasets are empty for the baseline. It is also worth noting that  some of the centroids are aligned with the data samples for $\minimumCardinality=0$. This is because the corresponding partitions have only one data sample. This implies that the federated $k$-means algorithm requires any extra precautions  such as noise injection for enhancing privacy \cite{li2022secure}. These issues are addressed for $\minimumCardinality=5$. In this case, the centroids are more localized in densely populated areas, resulting in a better representation of the users' locations. The centroids are likely not to be aligned with a specific user location as a partition has at least 5 data samples in this case.

\section{Concluding Remarks}
In this study, we propose a wireless federated $k$-means clustering algorithm along with an OAC scheme that does not require \ac{CSI} at the ES and EDs to address per-round communication latency. By considering data heterogeneity, we utilize a maximum-value adaptation method to reduce quantization error and a re-initialization strategy for a centroid that has a small cardinality in the corresponding partition to improve the performance of the algorithm. For a customer-location clustering scenario, we assess the proposed algorithm under different channel conditions and OAC configurations. Our results indicate that the proposed approach can perform similarly to the standard $k$-means while reducing the per-round communication latency notably. Future work will analyze the convergence of the proposed approach.


\bibliographystyle{IEEEtran}
\bibliography{references}

\end{document}